# Investigation of positive and negative modes of nanosecond pulsed discharge in water and electrostriction model of initiation

Yohan Seepersad[1,2], Mikhail Pekker[1], Mikhail N. Shneider[3], Alexander Fridman[1,4], Danil Dobrynin[1]

[1] A. J. Drexel Plasma Institute, Drexel University, Camden NJ 08103
[2] Electrical and Computer Engineering Department, Drexel University, Philadelphia PA 19104
[3] Mechanical and Aerospace Engineering Department, Princeton University, Princeton, NJ 08544
[4] Mechanical Engineering and Mechanics Department, Drexel University, Philadelphia PA 19104

## Abstract

This work investigates the development of nanosecond pulsed discharges in water ignited with the application of both positive and negative polarity pulses to submerged pin to plane electrodes. Optical diagnostics are used to study two main aspects of these discharges: the initiation phase, and the development phase. Nanosecond pulses up to $24\ kV$ with $4\ ns$ rise time, $10\ ns$ duration and $5\ ns$ fall time are used to ignite discharges in a $1.5\ mm$ gap between a copper plate and a tungsten needle with radius of curvature of $25\ \mu m$. Fast ICCD imaging is used to trace the discharge development over varying applied pulse amplitudes for both positively and negatively applied pulses to the pin electrode. The discharge is found to progress similar to that of discharges in long gaps – long sparks - in gases, both in structure and development. The more important initiation phase is investigated via Schlieren transmission imaging. The region near the tip of the electrode is investigated for slightly under-breakdown conditions, and changes in the liquid's refractive index and density are observed over the duration of the applied pulse. An attempt to explain the results is made based on the electrostriction model of discharge initiation.

## 1 Introduction

The topic of high-voltage interaction with liquids has held significant interest amongst the scientific community for a number of years, as per the work cited in the reviews [1-4]. In particular, [4] provides a more recent account of some of the applications employing plasma discharges in liquids with emphasis on its impact on the field of nanoscience. Much attention has also been placed on trying to identify the fundamental processes leading to the formation of plasma in liquid media. Until recently, the favored theory was that plasma generation in liquids was enabled via the formation of bubbles near the electrodes produced by various effects (Joule heating, electrostatic expansion of pre-existing micro-bubbles, electrochemical effects); these low-density cavities effectively facilitate electron avalanche sufficient to initiate breakdown [1, 3]. More intriguing work in the last couple of years has shown the possibility of plasma formation in liquids without the initial bubble growth stage [5, 6]. Plasma generation in water, produced by applied electric fields lower than the predicted thresholds for the direct ionization of condensed media was shown, and some attempt has been made to explain the mechanisms leading to these observations on a fundamental level [6-9]. Paramount to the conjectured mechanisms in these publications is the requirement of applied electric fields with extremely fast voltage rise times ($< 5\ ns$) – much like the conditions under which direct plasma formation in liquid phase without bubbles was observed in [5].

The work reported in [7-9] proposes that highly non-uniform electric fields with fast rising voltage fronts lead to the formation of nanoscopic voids near the pin electrode, facilitating rapid electron avalanches. Voltage pulses applied to a pin to plane electrode geometry generate a highly non-uniform electric field with a very strong gradient near the tip of the needle electrode. Dielectrics stressed with such strong non-uniform electric fields are affected by electrostrictive forces which cause deformation toward the region of higher electric field. In [7, 8] the authors argue that these electrostrictive effects can lead to the formation of a region near the electrode tip saturated by nanoscopic pores or voids. These predicted "voids" are dissimilar to bubbles in the sense that they represent liquid-ruptures rather than gaseous cavities (bubbles), and are of significantly smaller size. Furthermore, these results further lead to the hypothesis that these voids are not spherical, but rather elongated along the direction of the electric field lines [10]. These voids provide sufficiently long pathways within which electrons can gain sufficient energy from the applied electric field to cause ionization. The rapidly rising voltage front is necessary so that the electrostriction conditions dominate before hydrodynamic forces in the liquid have time to react to counteract the region of negative pressure associated with these voids [10]. The basic schematic depicting this initiation mechanism is shown in Figure 1. The predicted behavior proceeds independent of voltage polarity.

In this paper we analyze the development of nanosecond pulsed plasma in water for negative and positive polarity, on the basis of this electrostriction mechanism as a possible initiation phase. High voltage pulses with nanosecond rise times are used to ignite discharges in pure water using both positive and negative pulses applied to a pin to plane electrode setup, and minimum electric field threshold is estimated from the experimental results. ICCD imaging is used to study the emission phase of the discharge over the profile of the applied voltage pulse, and comparison is drawn between initiation and development of both positive and negative modes. For applied voltages slightly below the threshold required for plasma formation, transmission imaging is used to investigate changes in the liquid optical properties as a consequence of these nanosecond pulses to investigate the plausibility of the electrostriction models proposed in [6, 7, 10].

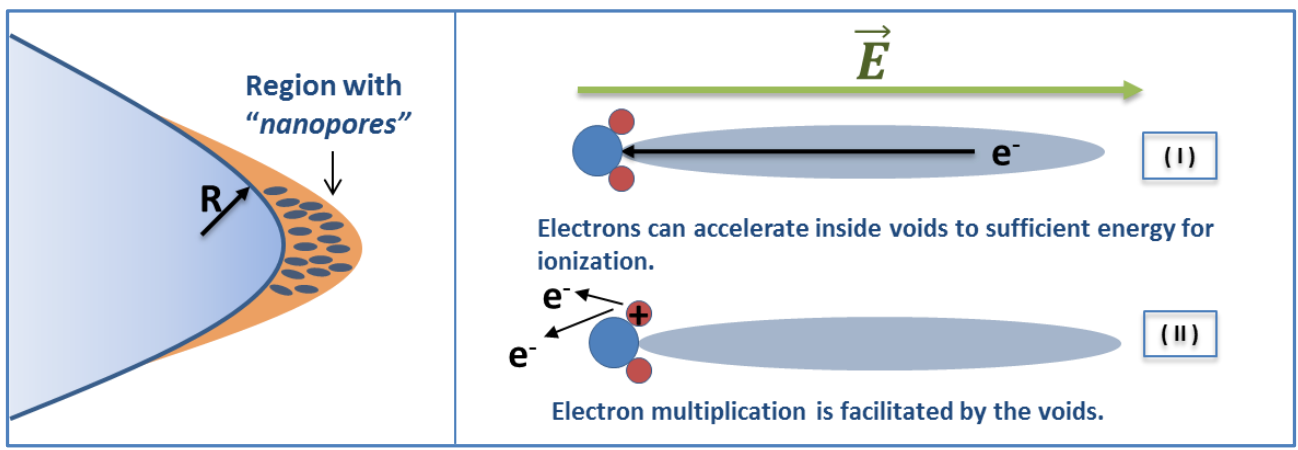

Figure 1: Basis of electrostriction mechanism for plasma initiation in water by nanosecond pulses in a pin to plane electrode geometry. R is the radius of curvature of the electric field, and E is direction of the electric field.

## 2  Experimental Setup and Methodology

The basic experimental setup utilized in these experiments mimics the ones used in [6, 8, 9] and in the simulations performed in [7]; the schematic is shown in Figure 2. Pulses with adjustable amplitude

($16.7–23.1 kV$) are generated by a nanosecond pulse generator from FID Technology (FID, GmBH). Voltage rise time was measured at $3 ns$, with $10 ns$ pulse duration at 90% amplitude and $4 ns$ fall time, and pulses were delivered to the electrodes via $100 ft$ of $RG\ 393/U$ high voltage coaxial cable (Figure 3 a). The lengthy cable would have introduced some degree of signal filtering which was estimated by analyzing the change in characteristics reflected pulse and assuming a linear change in amplitude and rise/fall rate per unit length of the cable. The effective amplitude change was $< 5\%$ and the estimated voltage rise/fall time at the end of the cable was $4ns\ /\ 5ns$ respectively. A return current shunt was mounted $22\ ft$ from output of the power supply and was used for pulse monitoring as well as camera signaling- this removed the uncertainty in triggering due to jitter observed in the power supply output ($\sim 5ns$). The shunt comprised 10 low inductance $3\Omega$ resistors soldered into a gap cut in the shield of the cable. The line current is then calculated as follows:

$$I^+_{line}(z,t) = \frac{Z_{out} + Z_{scope}}{Z_{scope}} \cdot \frac{V_{shunt}}{Z_{shunt}} \times \gamma \qquad (1)$$

Where: $I^+(z,t)$ is the line current, $Z_{out}$ is the output impedance of shunt connector ($50\Omega$), $Z_{scope}$ is the input impedance of the oscilloscope ($1M\Omega$), $Z_{shunt}$ is the shunt impedance ($0.3\Omega$), $V_{shunt}$ is the voltage measured across the shunt, and $\gamma$ is an attenuation factor from three $10dB$ attenuators used to step down the signal amplitude ($10^{3/2}$). Considering that $Z_{scope} \gg Z_{out}$ the equation (1) simplifies to:

$$I^+_{line}(z,t) = 105.41 \times V_{shunt} \qquad (2)$$

Considering that the BSC is a low energy tap of the signal on the line, and using a line impedance of $50\Omega$, the voltage pulse amplitude was calculated as:

$$V^+(z,t) = I^+_{line}(z,t) \times 50 \qquad (3)$$

Equation (2) and (3) were used to correct the amplitudes in the graph shown in Figure 3 (a). These calculations were verified to be fairly accurate by measuring the open circuit signal at the end of the line using a high voltage probe (Tektronix, P6015A), the measurements shown in Figure 3 (b). While the probe's frequency response limited its ability to accurately detect the pulse shape, the amplitude measurement was assumed accurate. Signal attenuation due to dielectric losses also account for the lower measured voltage amplitude under open circuit conditions. The lengthy cable also provided isolation of the twice reflected pulse from the electrodes with $292 ns$ delay.

The electrodes comprised a tungsten rod, mechanically sharpened to a radius of curvature of $25\mu m \pm 2 \mu m$, and a $18mm$ diameter copper plate, mounted in a pin-to-plane geometry. All experiments were performed with the electrodes submerged in distilled deionized water (type $II$) with maximum conductivity $1.0\ \mu S.cm^{-1}$ (EMD, Chemicals). Imaging was facilitated by the 4-Picos camera (Stanford Computer Optics), having a minimum gate time of $200 ps$, a spectral sensitivity in the range $250–780 nm$.

Synchronization of the images captured with the voltage pulse profile was done by careful consideration of all the delays involved in the system. Open-circuit pulse reflectometry measurements were used to determine the delay from the BCS signal to the camera as $114 ns\ \pm 0.2 ns$. The internal camera trigger propagation delay was specified by the manufacturer was $65.4 ns \pm 20 ps$. The signal delay from the BCS to the camera trigger input was measured as $23.4 ns$. The camera detector was $\sim 1 ft$ away from the discharge gap, and a propagation time of $1 ns$ was assumed for the optical delay between discharge emission and signal impinging on the MCP (micro channel plate). Minimum gating time used for images was $1 ns$, well above the cumulative uncertainty associated with the delays of the other components. Gate signaling was then controlled by carefully adjusting software imposed delays. We remark that image sequences constructed in these results relied on assumed good reproducibility of the phenomenon, as well as the ability to reliably capture repeated images at precisely the same moment in time (relative to the start of the voltage pulse).

The experimental setup was slightly modified to perform schlieren imaging. A $35 mW$ laser diode at $405 nm$ (D6-7-405-35-M, Egismos) was used to produce a collimated light source through the region near the electrode tip as shown in the exaggerated schematic in Figure 4. A microscope focusing lens was used to focus the beam, with the conjugate focal plane adjusted to coincide with the tip of the electrode (distance $p$ in Figure 4). A razor blade placed at the focal point of the lens (distance $f$ away in Figure 4) acted as the "schlieren stop". The 4-picos camera with a second focusing lens was used to capture the transmitted light, with similar considerations for various signal delays taken into account in order to determine the imaging-voltage synchronization. Assuming a Cartesian reference coordinate system as depicted in Figure 4, this setup allowed contrast imaging of refractive index ($\eta$) perturbations in the $xy$ plane, with intensity sensitivity to the first spatial derivative of $\eta$ ($\partial \eta / \partial x$ and $\partial \eta / \partial y$) occurring near the conjugate focal plane of the decollimating lens [11].

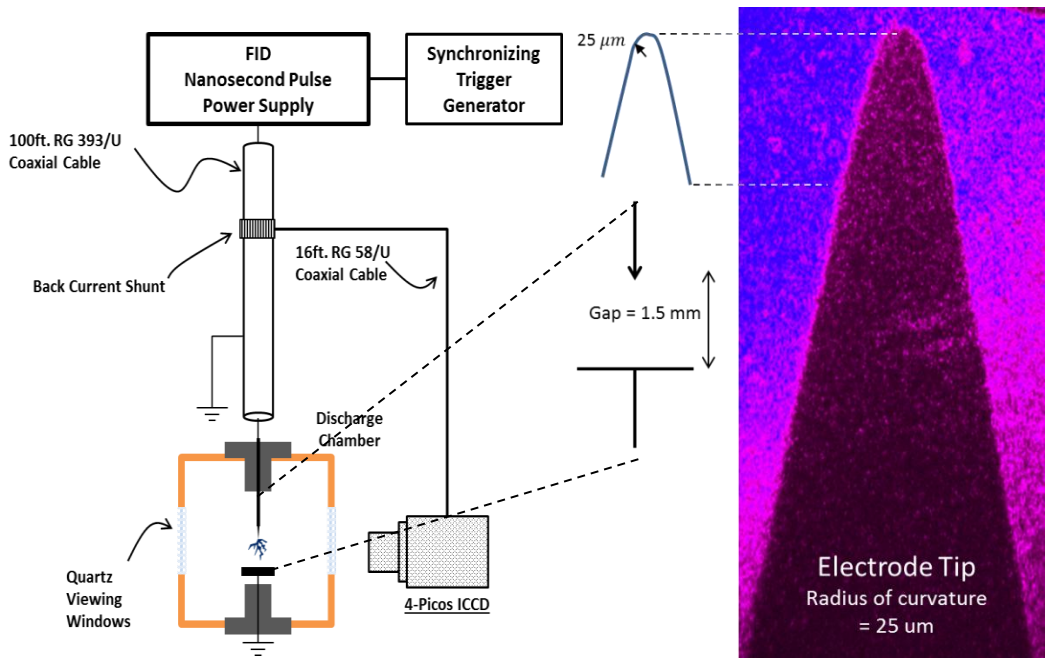

Figure 2: Basic experimental setup and laser-back illuminated image of electrode tip showing electrode profile

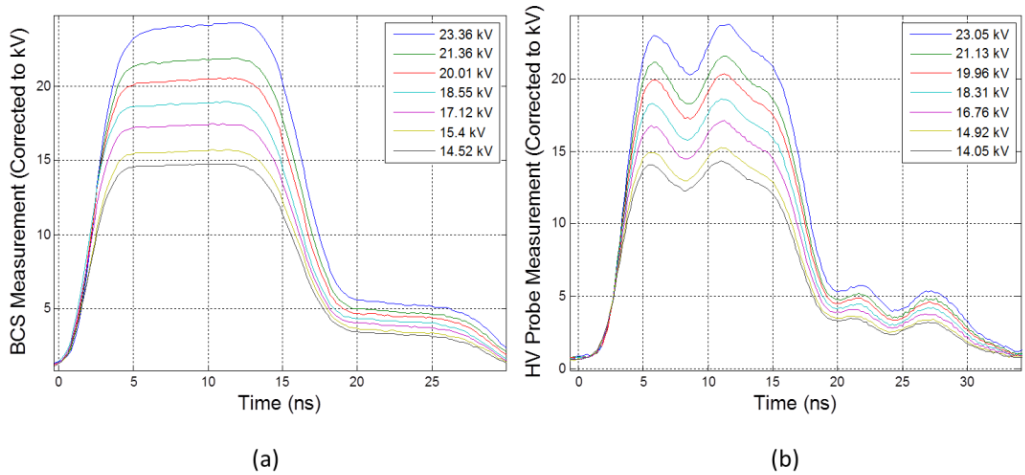

**Figure 3: (a)** Back current shunt measurements of applied nanosecond pulses used in experiments. Amplitudes shown are calculated based the estimation of the line current from the voltage drop across the known shunt impedance.
**(b)** Open circuit measurements by high voltage probe.

The schlieren images taken were all post processed using background-subtraction, the background image captured with no voltage applied to the electrodes, and a typical image is shown in Figure 5. A diffuse glow or halo outlines the profile of the electrode, and is a result of diffraction of light around the edges of opaque objects commonly observed in schlieren systems [11].

Changes in refractive index occurring in conjugate focal plane of the schlieren setup can manifest themselves in 2 ways. Light rays through these regions will be bent off the $z$ axis in some direction normal to the $xy$ plane by some small angle. Deflections occurring in the plane parallel to the knife edge will not be seen in this setup. Deflections occurring in the plane perpendicular to the knife edge can either deflect rays into the knife edge (dashed blue line) or away from the knife edge (dashed red line). Thus, in the imaging plane, brighter areas will correspond to refracted light that has been displaced above the knife edge. It is worth noting that light deflected into the knife edge (dashed blue line) will appear as a darker region, relative to what the background image would look like for homogeneous media. This is because the spot size of the condensed beam at the focal point has a finite size [11]. This explains why the background image shown in Figure 5 is not completely dark in the areas besides the diffraction halo. Furthermore, considering that light bends towards regions of higher refractive index and away from regions of lower refractive index, proper positioning of the knife edge will allow regions of the higher density and regions of lower density to be distinguished using a background subtraction scheme.

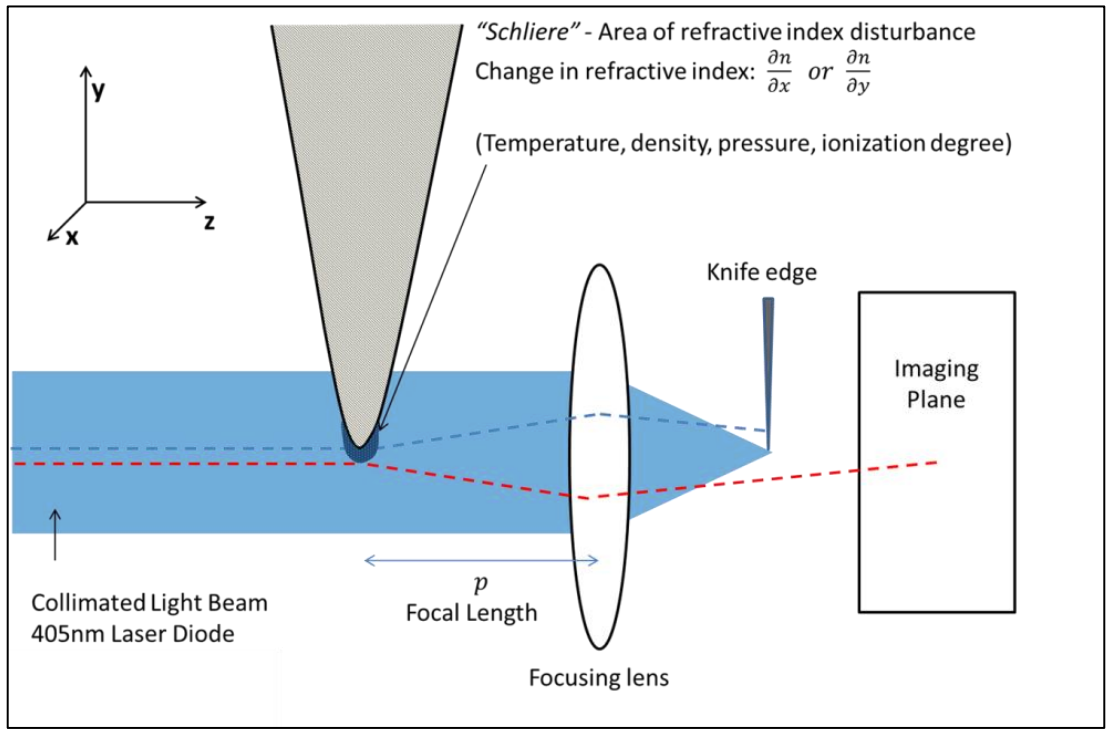

Figure 4: Schematic showing the principle of operation of the basic laser schlieren setup.

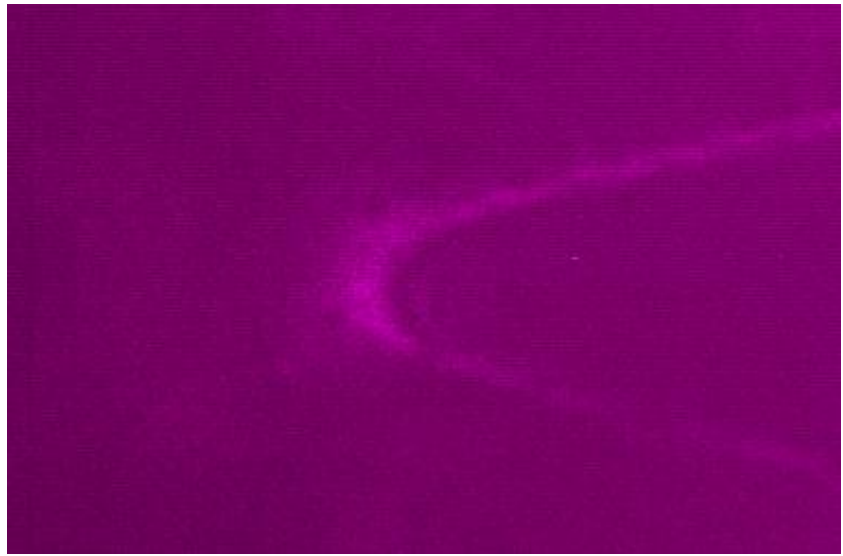

Figure 5: Background image used to correct schlieren images. The outline in the image shows the diffraction halo commonly observed around opaque objects in schlieren systems.

# 3 Results

The results presented in this section are contains mostly ICCD images and image-analysis based on the different experiments concerned. The images have been assigned an artificial color map which represents the intensity distribution of light over the image. Brighter areas represent regions of intense incident light in that region, and darker areas the lack thereof.

## 3.1 Breakdown Conditions: ICCD Imaging

Previous studies have been done investigating the development of nanosecond pulses in liquids [5, 6]. The results included here are consistent with the findings of the previous works, but further delves into the behavior of the discharge development as the voltage pulse amplitude is varied. Three different regions throughout the lifetime of the applied pulse over the electrode gap are considered: the first $4ns$ corresponding to the rising edge of the pulse; the following $10ns$ corresponding to the voltage plateau; and, the final stage comprising the falling edge ($5ns$) and the transient remnant voltage which was an artifact of the applied pulse not completely going to zero after the main falling edge (see Figure 3). Positive and negative pulses with amplitudes $V = 23.1 kV$, $21.1\ kV$ and $20.0\ kV$, were applied to the electrodes (HV to pin). Under these conditions, the voltage amplitude was sufficient to ignite discharges in the liquid, the evolution of which is shown Figure 6, Figure 7, and Figure 8 for positive pulses and Figure 9 for negative pulses.

### 3.1.1 Positive Mode

From these pictures, several observations regarding the size of the plasma region formed versus the applied amplitude can be made. Firstly, we remark that time at which the discharge emission begins increases at the voltage amplitude decreases (see Figure 6). The rise time for all the pulses is the same ($4ns$), thus $dV/dt$ is different for each voltage setting. The electric field intensity near the tip when the first emission is observed can be calculated from:

$$E = \frac{2V(t)}{r \ln\left(\frac{4d}{r}\right)} \qquad (4)$$

Where: $V(t)$ is the voltage across the gap at the time, $t$, calculated from $V(t) = t \times V_{max}/\tau_{rise}$, were $V_{max}$ is the peak amplitude, $\tau_{rise} = 4ns$. Also, $r = 25 \mu m$ and $d = 1.5mm$ are the radius of curvature of the needle and electrode gap respectively. Considering that the gating time for the camera was $1ns$, we estimate rather the range of the electric field intensity that exists over the gating window for which emission was first observed. For instance, the first emission observed at time $t = 1ns$ for an applied voltage of $23.1\ kV$ as seen in Figure 6 represents a gating window over the pulse from $t = 1ns$ to $t = 2ns$, the times relevant to the voltage pulse shown in Figure 3 (a). We found that for pulse amplitude of $23.1\ kV$, first emission appeared when the electric field was $0.84 - 1.69\ MV\ cm^{-1}$, for pulse amplitude of $21.1\ kV$ the field was $1.54 - 2.31\ MV\ cm^{-1}$, and for a pulse amplitude of $20.0\ kV$ and field was $2.19 - 2.92\ MV\ cm^{-1}$. Thus, as $dV/dt$ increased, the threshold electric field at which emission was first seen decreased. This explains why no discharge was seen for applied voltages of $18.3\ kV$ where the maximum electric field on the tip was $2.7\ MV\ cm^{-1}$ but the rate of change of voltage increase was lower.

Secondly, the size of the emitting region grows only with the rising edge of the pulse, with a maximum size attained at the transition between the rising edge of the pulse and the voltage plateau (compare Figure 6 at 4ns and Figure 7 at 5ns). When $dV/dt$ fell to zero throughout the middle of the pulse, a distinct phase where the plasma was extinguished was clearly apparent. This is consistent with the dark phase observed in [5, 6], although the emission does not completely disappear immediately. Rather, the appearance changes to a more filamentary structure ($9ns - 13ns$), eventually diminishing to a small glow ($14ns$).

Finally, the emitting region size grows on the falling edge of the pulse, with a peak size occurring at $18ns$ (see Figure 8). In the final $8ns$ of the discharge lifetime, the plasma extinguished uniformly, in a ball-like structure, rather than the observed filamentary disappearance when the voltage amplitude was high.

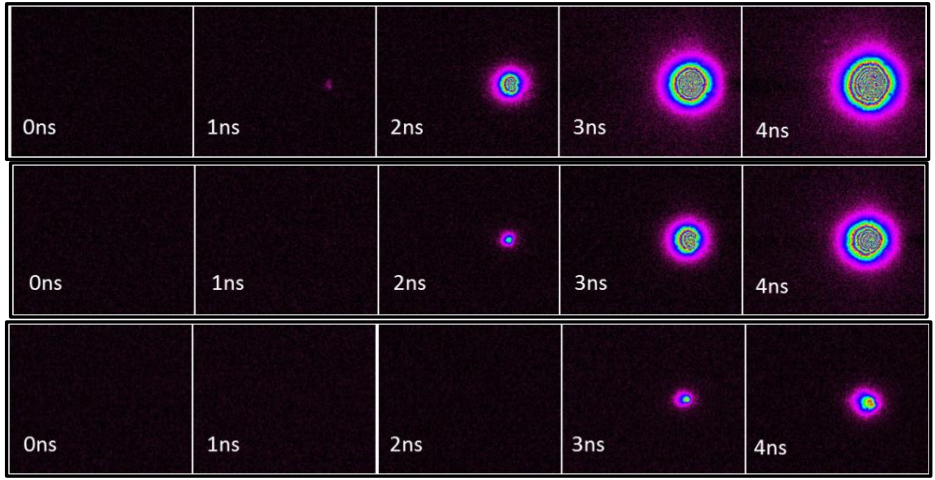

**Figure 6:** Development of initial stage of discharge in water for varying positively applied voltages to the pin electrode. These first 4ns is the effective time over which the rising edge of the voltage pulse appears on the electrodes.
Top Row: +23.1kV; Center Row: +21.1 kV; Bottom Row: +20.0 kV.
All images are 7.6 x 6.4mm, with 50 accumulations captured over 1ns exposure, and each image is shifted in time by 1ns.

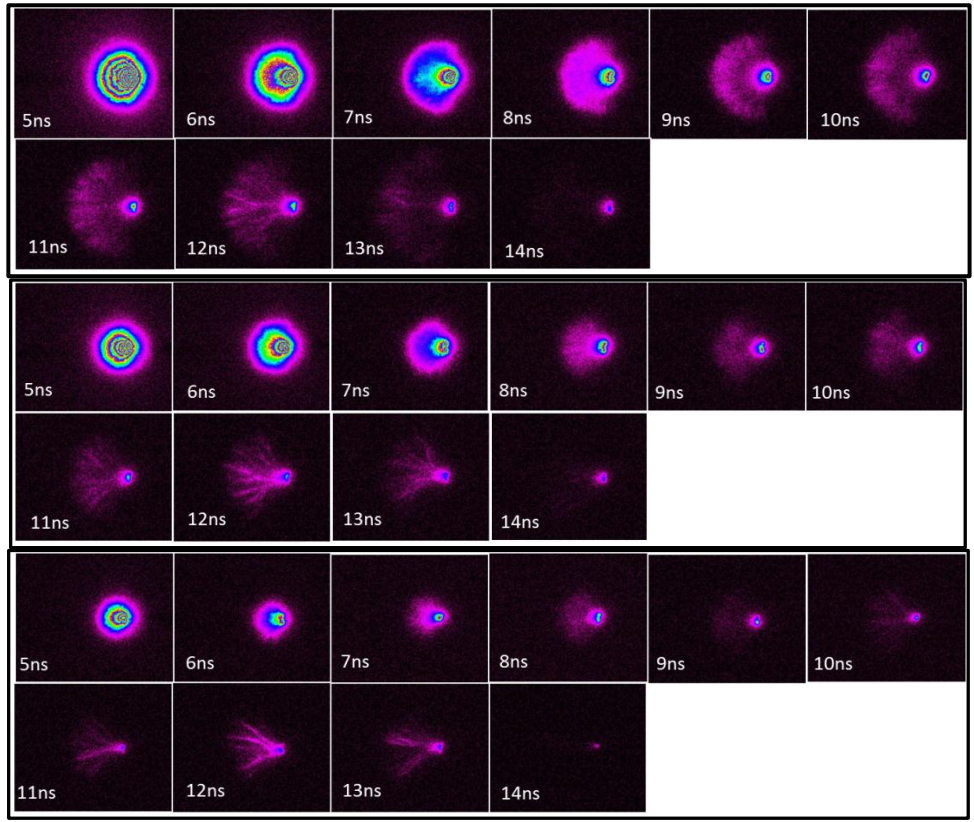

**Figure 7:** Development of discharge in water for varying positively applied voltages to the pin electrode over the plateau of the applied pulse. These 10ns are the effective time over which the voltage is constant over the electrode gap.
Top two rows: +23.1 kV; Center 2 rows: +21.1 kV; Bottom 2 rows: +20.0 kV.
All images are 7.6 x 6.4mm, with 50 accumulations captured over 1ns exposure, and each image is shifted in time by 1ns.

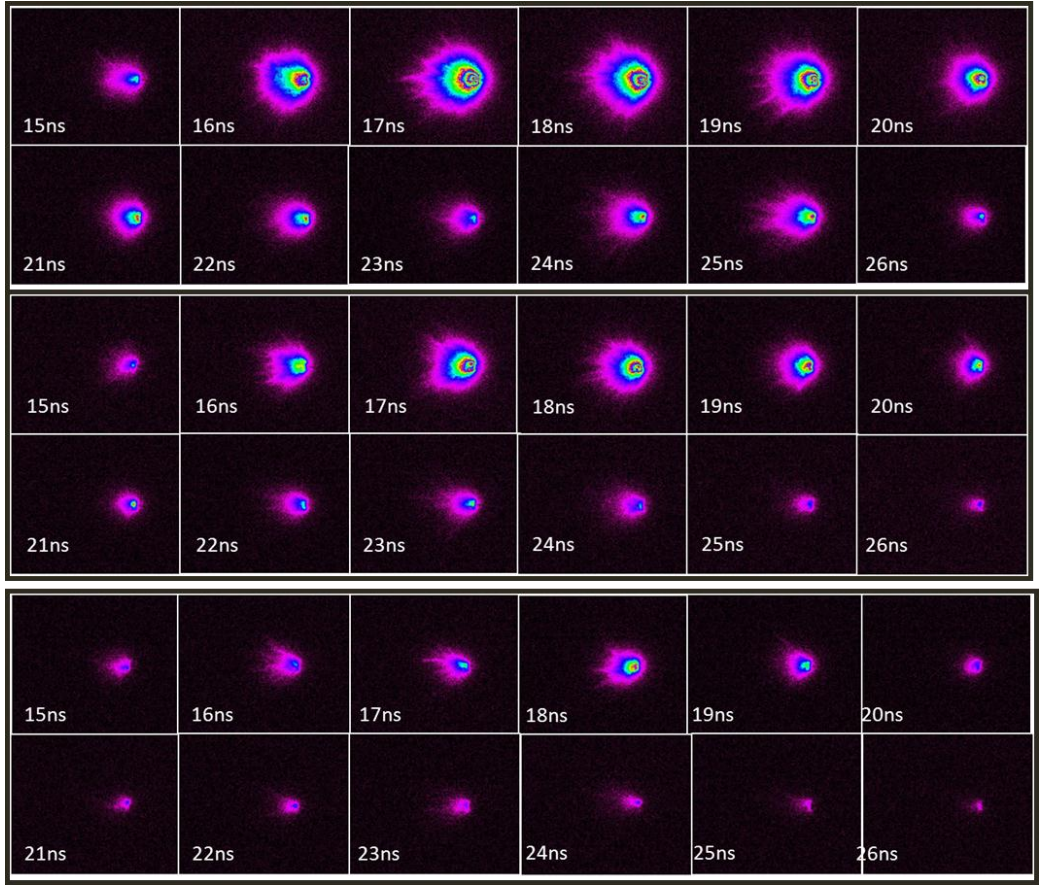

**Figure 8: Development of discharge in water for varying positively applied voltages to the pin electrode over the falling edge of the applied pulse. These images from 15ns to 20ns ns are taken over the effective time of the falling edge of the pulse across the electrode gap.
Top two rows: +23.1 kV; Center 2 rows: +21.1 kV; Bottom 2 rows: +20.0 kV.
All images are 7.6 x 6.4mm, with 50 accumulations captured over 1ns exposure, and each image is shifted in time by 1ns.**

### 3.1.2 Negative Mode

Work involving nanosecond discharges in water with negatively applied pulses to the needle in a pin-to-plane electrode configuration has also been performed in [12], although the development of the discharge was not reported on. Similar time-resolved imaging development as reported on for the positive mode were conducted for negatively applied pulses to the pin electrode, with amplitudes $-23.1 kV, -21.2 kV$ and $-20.0 kV$.

First emission always occurs $2ns$ after the start of the pulse, and the estimated electric field intensity ranges at which the emission is first seen can be estimated as done for the positive mode. For an applied pulse of $-23.1\ kV$ the first emission was seen when the electric field near the tip was $1.7 - 2.5\ MV\ cm^{-1}$; for a pulse of $-21.1\ kV$ the field was $1.5 - 2.31\ MV\ cm^{-1}$; and, for a pulse of $-20.0\ kV$ the field was $1.5 - 2.2\ MV\ cm^{-1}$. As was the case for positive mode of discharge, no emission was seen when a pulse of $-18.3\ kV$ was applied.

In stark contrast to the emission size and patterns seen in the positive mode, the discharge in the negative mode appears as only a faint glowing ball, the size of which is not larger than the diameter of the electrode tip ($\sim 45 - 50 \mu m$), as shown in Figure 9. The faint emission appears to start at the same time ($2ns$) for all applied voltages, and the maximum size of the emitting region corresponds to the time where the rising edge transitions to the voltage plateau. It should be noted that the size of the emitting region also relatively constant, with no appreciable variation as observed for positive pulses. Similar to the positive mode, no emission is seen when $dV/dt$ falls to zero and the emission phase only reappears on the falling edge of the pulse.

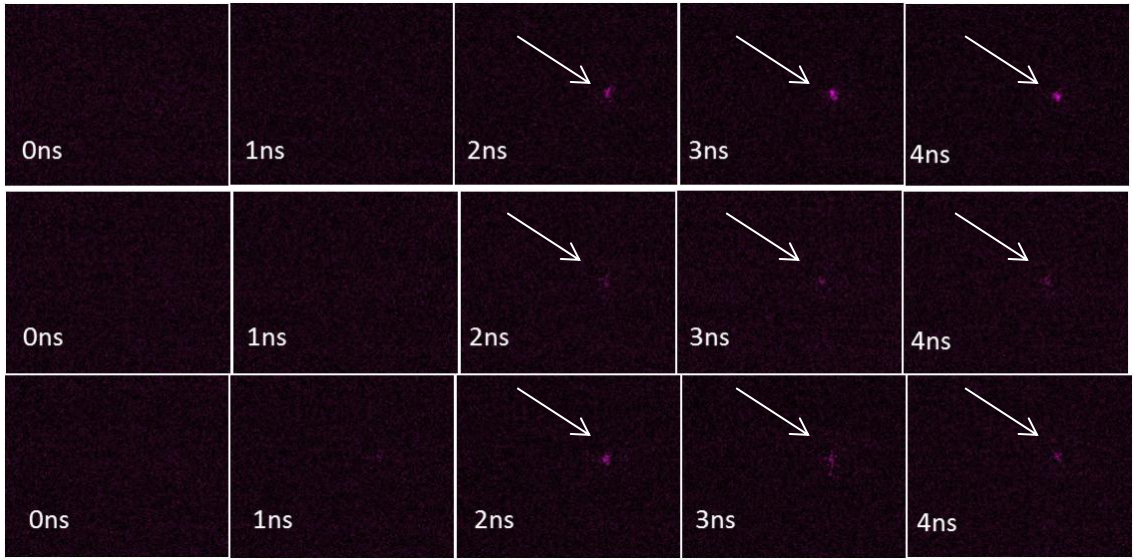

Figure 9: Development of initial stage of discharge in water for varying negatively applied voltages to the pin electrode. These first 4ns is the effective time over which the rising edge of the voltage pulse appears on the electrodes.
Top Row: -23.1kV; Center Row: -21.1 kV; Bottom Row: -20.0 kV.
All images are 7.6 x 6.4mm, with 50 accumulations captured over 1ns exposure, and each image is shifted in time by 1ns.

## 3.2 Under-breakdown Conditions: Schlieren Imaging

For applied voltages which initiate breakdown, discharge emission was too intense to allow proper utilization of Schlieren imaging for observing density changes during the formation of plasma. Thus, the liquid surrounding the electrode tip was studied for under-voltage conditions, not sufficient for plasma to be initiated.

Figure 10 shows the schlieren imaging results for the initial stage of an applied pulse voltage of $+16.8\ kV$. The label $-1ns$ on the first picture annotates what the region near the tip of the electrode is imaged as, $1ns\ before$ the pulse is applied to the electrodes. We notice that the only changes in the transmission characteristics of the liquid near the electrode tip must be as a result of the applied voltage. Since this picture is background-subtraction processed, the uniformly colored image confirms that there is no difference between the background image (with no pulse applied), and the image immediately before the pulse reaches the electrode. When the voltage beings to rise on the electrode gap however the results show the growth of a dark region near electrode tip. Considering the orientation of the knife edge in the schlieren setup, this can be interpreted in two ways. Firstly, this dark region could be seen as an area of reduced refractive index from the surrounding liquid. This would lead to light refraction into the path of the knife edge, and as a result a region of lower light intensity would appear in the image plane. Another interpretation is that a dispersive region which scatters light is formed, reducing the transmitted light to the image plane. The exact mechanism does not distract form the observation that some distinct change in the optical properties of the liquid is taking place as a result of the application of these pulses. Moreover, these changes manifest themselves on a nanosecond timescale

Similar results also exists for negatively applied voltages, as shown in Figure 11 where a pulse of $-18.3 kV$ was applied to the pin electrode. Variations in intensities and clarity between Figure 10 and Figure 11 were due to slight changes made to the alignment of the optics when switching between experiments. The behavior of the fluid near the electrode upon application of the pulse was essentially the same however.

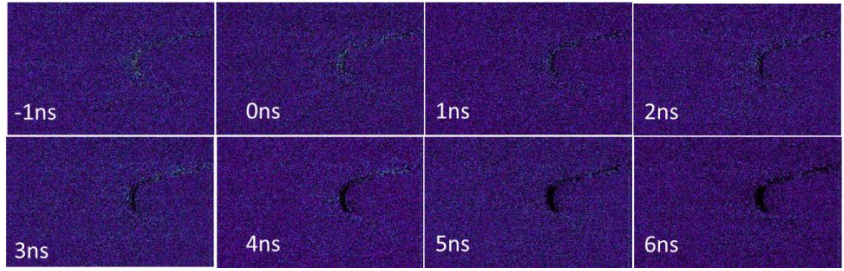

**Figure 10: Time resolved Schlieren images taken for a +16.8 kV pulse. Images are 365 x 233um, 100 accumulations and 1ns step and 1ns camera exposure.**

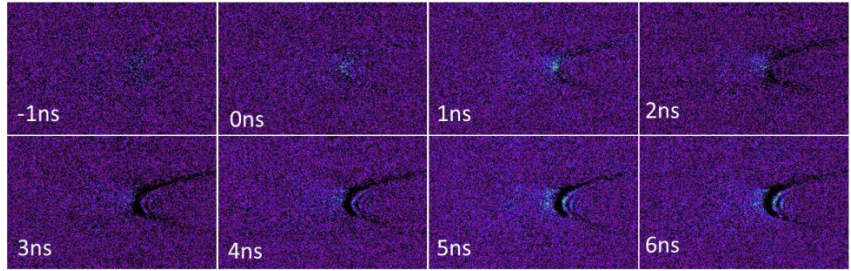

**Figure 11: Time resolved Schlieren images taken for a -18.3 kV pulse. Images are 365 x 233um, 100 accumulations and 1ns step and 1ns camera exposure.**

# 4 Discussion

In this section we will address the development of the discharge in both the positive and negative modes in the first section, assuming that initiation has already occurred. The second section will discuss the possible mechanism of initiation which we propose occurs independently of the voltage polarity.

## 4.1 Positive and Negative Mode of Discharge Development

These results show that plasma generated by nanosecond pulses exhibit behavior very similar to that of discharges in long gaps – long sparks - in gases, both in structure and development. In the case of microsecond pulses reported in [13], the authors showed that both positive and negative pulses produced plasma with branching filamentary structure when the initiation mechanism was clearly identified as being related to bubble growth at the electrode tip. In these experiments however, filamentary discharge structure only exists for positively applied pulses. Positive discharge initiation is associated with cathode directed streamer propagation and is governed by the generalized Meek criterion for breakdown [14]:

$$\int_0^{x_{max}}[\alpha(x) - \beta(x)]dx \geq 20 \qquad (4)$$

On the other hand, negative discharge initiation is governed by secondary emission from the cathode, and discharge develops similar to a Townsend breakdown mechanism governed by the criteria [14]:

$$\int_0^{x_{max}}[\alpha(x) - \beta(x)]dx = \ln\left(1 + \frac{1}{\gamma}\right) \qquad (5)$$

Where: $\alpha, \beta$ and $\gamma$ are the first, second and third Townsend coefficients respectively. In these equations, $x_{max}$ represents the length where the ionization rate and recombination rate are equal, or $\alpha(x_{max}) = \beta(x_{max})$, and corresponds to the maximum size of the visible glowing region for negative corona.

Positive discharge develops via a leader mechanism where the local electric field in the head of the streamer drives its propagation, thus it is possible for the discharge to grow much larger the initial region where the primary avalanche takes place. A continuously rising electric field sustains propagation of the leader and discharge growth which halts when $dV/dt$ falls to zero: as the leader propagates further away from the electrode and positive charge of the discharge column increases, a reverse radial electric field appears in order to compensate this excessive positive charge by radial electron current. This decreases axial electron current in the channel, resulting in decrease of the leader head potential leading to extinguishment of the discharge. Thus, as in the case of leader discharge development in long gaps, constant increase of electric field is essential to the discharge development [15]. On the falling edge of the pulse, the electric field as a result of the positive space charge generated during the first strike is strong enough to re-ignite the plasma, thus the emitting region again grows on the falling edge of the pulse. Since the medium already contains pre-ionized channels from the initial streamers, the plasma reignites more easily. Positive mode development is pictorially represented in Figure 12.

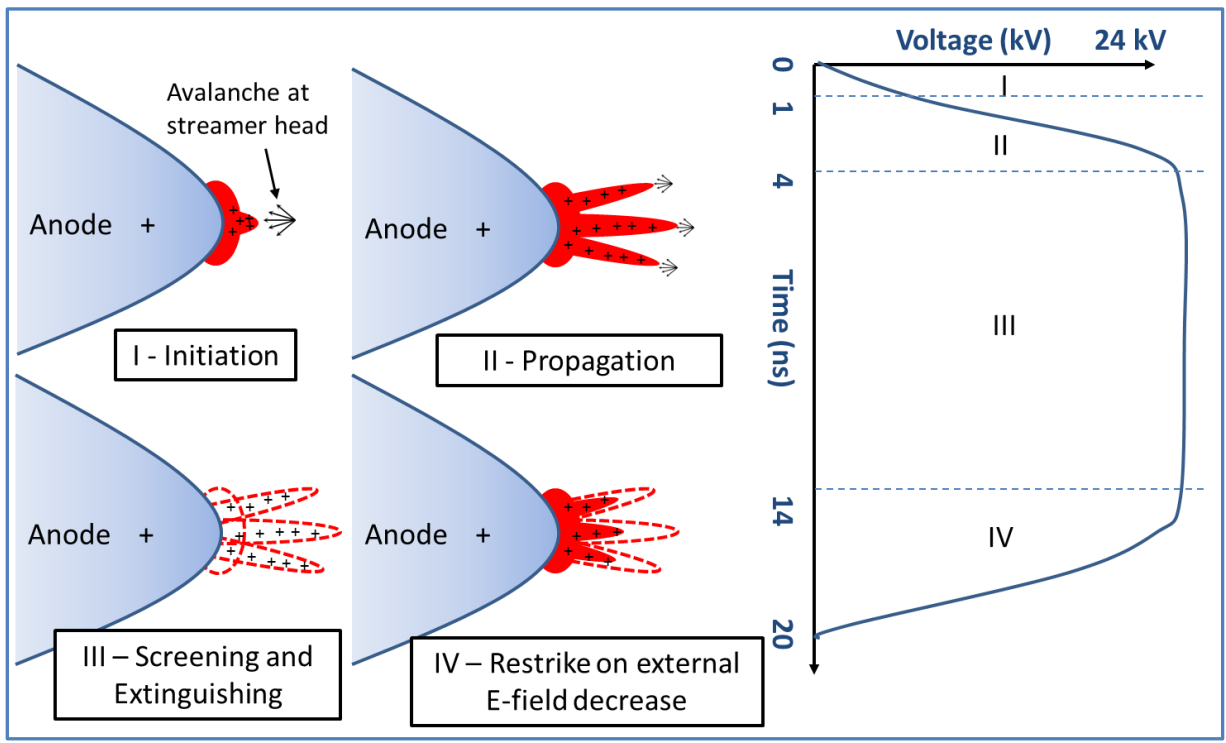

Figure 12: Development of positive mode nanosecond pulsed plasma in water: (I) Initiation, (II) propagation, (III) extinguishing on the voltage plateau and then (IV) reilluminations on the falling edge. Inset is the voltage pulse showing the relative times at which the phases occur.

The negative mode discharge results in weak corona-type glow appearance without leader formation and does not propagate since electrons are being repelled by the electrode and get rapidly solvated by water molecules further away from the electrode. Similar to gases, breakdown electric field in the case of negative polarity of the needle is greater than that of positive discharge and is related to the differences in the avalanche and streamer development conditions [15]. In the vicinity of needle anode, avalanches are directed towards the electrode and experience the effect of greater electric field as they approach it. In contrast, cathode-initiated avalanches are directed away from the electrode – in the direction of lower electric field. Secondary emission is able to sustain discharge near the region of the electrode tip where the electric field is highest and liberated electrons from collision processes and secondary emission can continue to participate in ionization. At a distance $x_{max}$ away from the electrode, attachment become significant, and ionization cannot continue. Solvated electrons which are trapped lead to electric field screening and subsequent extinguishing of the plasma when $dV/dt$ goes to zero on the voltage plateau. As with the positive mode, the plasma reignites on the falling edge, but again the structure is that of a relatively small corona glow. We have offered an explanation regarding the discharge development over time, as well as the major qualitative differences in structure and size between the positive and negative modes of discharge. The major question in these experiments still remains as to what facilitate the initiating electron avalanche in the liquid. The results from the Schlieren experiments however might offer some insight into this initial phase.

## 4.2 Discharge Initiation Mechanism

Changes in refractive index have been previously studied for nanosecond pulsed electric fields applied to submerged pin-to-plane electrodes in water, in which the authors attribute the well-known Kerr effect as the main process taking place [16, 17]. Diploe alignment results in changes in the permittivity of the water in the vicinity of the electrode. This dipole alignment with the electric field contributes to the ponderomotive forces which lead to dielectric deformation as discussed in [7-10] and references therein. The schlieren results presented in this report clearly depict some noticeable change in optical properties of the water layers close to the electrode tip where the electric field is highest, see: Figure 10 and Figure 11. This effect is insensitive to the polarity of the applied electric field, as the observed refractive index anisotropicity is seen for both positive and negative voltages. We remark that the appearance of the discharge emission under breakdown conditions coincides with the same time that these anisotropic regions are observed to be formed for under-breakdown conditions. It is quite obvious that the processes taking place in this region must lead to the initiation of the discharge in both modes. If we assume that the model presented in [7, 8, 10] holds true, then a possible explanation of the results will proceed in the manner which follows.

After $1-2ns$ we observe the formation of the region presumably saturated with nanopores for under-breakdown conditions, and we assume that at higher voltages, this region supports electron avalanche that initially forms a "seed" plasma region. This process should take place irrespective of the voltage polarity, since the electron acceleration occurs within the elongated voids produced near the tip. This could explain why the initial stages of the positive and negative modes of discharge appear the same (compare Figure 6, top row at $1ns$ and Figure 7, top row at $2ns$). We remark that these images represent a relatively coarse time evolution - $1ns$ exposures and $1ns$ steps. It is possible that the time of first emission is the same, occurring at some time very close to $2ns$ where the camera gating time could overlap. The change in the experimental setup from positive to negative pulses applied to the electrodes may have introduced some small signal path delay through the connectors to the electrodes which could also have introduced as much as $200ps$ delay between modes. Considering that, the main qualitative observation that the initial discharge shape was relatively the same suggests close similarity between the initial stages of both modes.

The comparisons drawn between these experimental results and the electrostriction model for liquid breakdown are based on the work reported mainly in [7, 10]. From [7] the predicted size of the region near the electrode within which the liquid can rupture is estimated by:

$$R < r_0 \left(\varepsilon\varepsilon_0 \frac{V^2}{P_c r_0^2}\right)^{1/4} = 0.75 \times 10^{-5} r_0^{1/2} V^{1/2} \quad (1)$$

where: $r_0$ is the radius of the tip of the electrode, $V$ is the applied voltage amplitude, $P_c$ is the critical negative pressure for water rupture, and $R$ is the maximum size of the region saturated with voids, and $\varepsilon$ and $\varepsilon_0$ are the permittivity constants of water and vacuum respectively. For the conditions in this paper, $r_0 = 5\mu m$ and

$V = 22.4 kV$ the estimated size of the region within which liquid rupture is expected is $5.8 \mu m$, assuming a critical value of negative pressure of $P_c = 20 MPa$ as in [7]. The size of the observed region of low density from Figure 10 can be estimated as being $5 - 10 \mu m$ thick, in good correlation with the model prediction.

For both positive and negative applied voltages to the tip, the formation of this dark region appears within the first $2 ns$ of the applied pulse. The authors in [7] formulate that the initial force acting on the liquid is proportional to the electric field squared, according to the equation:

$$\vec{F} \approx \frac{\alpha \varepsilon \varepsilon_0}{2} \vec{\nabla} E^2 \quad (2)$$

Where, $F$ is the body force acting on a polar fluid (water) and $\alpha$ is an empirical constant for polar liquids. This implies that the appearance of these "voids" should be independent of voltage polarity, which is the case in these experiments. The generation of these low density regions is also associated with a region of negative pressure which is generated much faster than the time needed for hydrodynamic forces to counteract. The model predicts the generation of this negative pressure zone upon the application of the voltage pulse, followed by a compression layer forming in a finite time afterward, as the hydrodynamic forces pushes the fluid toward the electrode due to unbalanced pressure. This compression layer later relaxes and propagates away from the electrode as the negative pressure goes to zero. Since the force is proportional to the amplitude of the electric field squared, the negative pressure that develops should be much stronger for pulses with higher amplitude. The propagating wave which follows is similar to the findings presented in [8].

We also observed that the plasma initiation somewhat depended on the voltage rise time of the applied pulse. The threshold electric field of first emission was higher for slower rising pulses. The only time-dependent relationship that can be drawn from this electrostriction mechanism is that relating the characteristic time of the reaction of the liquid's hydrodynamic forces to the negative pressure zones generated near the electrode. Faster rising pulses would produce voids faster, which can grow to critical size at a lower electric field. Slightly slower rising electric fields will allow hydrodynamic compression at lower electric fields, and thus it will take a longer time for critical void size to be created. This also explains why no discharge is seen for an applied voltage with $V_{max} = 18.3\ kV$ where the maximum electric field was higher than the threshold electric field when plasma was formed with faster rising pulses. We remark on how sensitive the phenomenon is to small changes on voltage rise rate: breakdown is initiated under $2.5\ MV\ cm^{-1}$ for a pulse with a front of $5.77\ kV\ s^{-1}$, but no discharge is seen for a pulse with a rising front of $4.5\ kV\ s^{-1}$ even with a maximum electric field of $2.7\ MV\ cm^{-1}$.

# 5 Conclusion

In this work we show the major qualitative differences existing between the positive and negative modes of nanosecond pulsed discharges in water, and that discharge development strongly resembles the manner in which positive and negative leader-type discharges behave in gases. The rate of voltage rise also affects the discharge initiation, since breakdown threshold for faster rising pulses is lower than for slower rising pulses. This could be linked to the rate at which the so called "nanopores" in the electrostriction mechanism are forming, and the ensuing rate of hydrodynamic pressure balance. Transmission imaging reveals the formation of a zone of altered optical refractive index on the same timescale as plasma is initiated. Undoubtedly this zone of anisotropicity is linked to the plasma initiation mechanism, and one possible explanation resides in the electrostriction mechanism presented in [7]. Further work studying the behavior of this liquid inhomogeneity could offer better insight into the nature of the change causing these changes in the liquid.

# Acknowledgment

This work was supported by Defense Advanced Research Projects Agency (grant # DARPA-BAA-11-31).